\documentclass[12pt,preprint]{aastex}

\newcommand{\be}{\begin{equation}}
\newcommand{\ee}{\end{equation}}
\newcommand{\ba}{\begin{eqnarray}}
\newcommand{\ea}{\end{eqnarray}}

\shorttitle{Direct Wavelet Expansion of the Primordial Power Spectrum}
\shortauthors{Mukherjee \& Wang}

\begin{document}

\title{Direct Wavelet Expansion of the Primordial Power Spectrum}
\author{Pia~Mukherjee$^{1}$, and Yun~Wang$^{1}$}
\altaffiltext{1}{Department of Physics \& Astronomy, Univ. of Oklahoma,
                 440 W Brooks St., Norman, OK 73019;
                 email: pia,wang@nhn.ou.edu}

\begin{abstract}
In order to constrain and possibly detect unusual physics during inflation, 
we allow the power spectrum of primordial matter density fluctuations,
$P_{in}(k)$, to be an arbitrary function in the estimation of
cosmological parameters from data. 
The multi-resolution and good localization properties of 
orthogonal wavelets
make them suitable for detecting features in $P_{in}(k)$.
We expand $P_{in}(k)$ directly in wavelet basis functions. 
The likelihood of the data is thus a function of the wavelet coefficients 
of $P_{in}(k)$, as well as the Hubble constant $H_0$, 
baryon density $\Omega_b h^2$, cold dark matter density $\Omega_c h^2$ and 
the reionization optical depth $\tau_{ri} $, in a flat $\Lambda$CDM cosmology.
  We derive constraints on these parameters from 
Cosmic Microwave Background (CMB) anisotropy data
(WMAP, CBI, and ACBAR) and
 large scale structure (LSS) data (2dFGRS and PSCZ) using the Markov Chain 
Monte Carlo (MCMC) technique.
The direct wavelet expansion method is different and complimentary to 
the wavelet band power method of Mukherjee \& Wang (2003a,b), and
 results from the two methods are consistent.
In addition, as we demonstrate, the direct wavelet expansion method 
has the advantage that once the wavelet coefficients have 
been constrained, the reconstruction of $P_{in}(k)$ can be effectively
denoised, i.e., $P_{in}(k)$ can be reconstructed using only the 
coefficients that, say, deviate from zero at greater than 1$\sigma$.
In doing so the essential properties of $P_{in}(k)$ are retained. The
 reconstruction also suffers much less from the correlated errors of 
binning methods.
The shape of the primordial power spectrum, as reconstructed 
in detail here, reveals an interesting new feature 
at $0.001 \la k/\mbox{Mpc}^{-1} \la 0.005$. It will be interesting
 to see if this feature is confirmed
by future data. The reconstructed and denoised $P_{in}(k)$ is favored over 
the scale-invariant and power-law forms at $\ga 1\sigma$.

\end{abstract}



\section{Introduction}

Inspite of the success of the inflationary paradigm
\citep{Guth81,Albrecht82,Gott82,Linde83,Kolb&Turner,Hu03,peeblesratra},
we are still in search of the correct model of inflation.
The simplest models of inflation predict a
power-law primordial matter power spectrum (for example, 
\cite{Linde83,naturalinf,extendedinf}).
However, there are many other viable models of inflation which 
predict primordial power spectra which cannot be parametrized 
by a simple power law 
(for example, 
\cite{Holman91ab,Wang94,Randall96,Adams97,Les97}).
These can represent unusual physics in the very early universe
(for example, see \cite{Chung00,Enqvist00,Lyth02,FengZhang03}), and then 
assuming a power-law primordial matter power spectrum
could lead to our missing the discovery of the possible features
in it  and erroneous estimates of the cosmological parameters
\citep{Kinney01}.

With results from WMAP, CMB data continue to be consistent with inflation
(Bennett et al. 2003, Spergel et al. 2003), having even detected 
the predicted anti-correlation between CMB temperature and
 polarization fluctuations near $l$ of 150. While the derived 
cosmological parameter constraints are consistent with previous 
constraints from independent and complementary observations, 
Peiris et al. (2003) report that the data, both with and without
 the use of complementary LSS data, seem to indicate a preferred
 scale in the primordial power spectrum. They conclude that the data
 suggest at the 2$\sigma$ level, but do not require, that the scalar
 spectral index runs from $n_s >1$ on large scales to $n_s < 1$ on 
small scales. Their analysis was done assuming a primordial power 
spectrum of power-law form about a pivot scale of 0.05 Mpc$^{-1}$
 (0.002 Mpc$^{-1}$) when using scalar (both scalar and
 tensor) modes.

Now that such precision measurements of CMB anisotropy
 have been made, we aim to allow the primordial power
 spectrum to be an arbitrary function, independent of any
 particular inflationary model, and use the data to place 
constraints on its shape. WMAP data underscore the importance 
of such measurements suggested earlier by \cite{Wang99, WangMathews02}
 who used conventional top-hat binning and linear interpolation
 respectively in the parametrization of $P_{in}(k)$.
Mukherjee \& Wang (2003a) parametrized $P_{in}(k)$ in terms 
of wavelet band powers, 
and arrived at a detailed measurement of $P_{in}(k)$ using pre-WMAP CMB data. 
A preliminary indication of a feature was reported. The method was shown
 to work better than the linear-interpolation method of \cite{WangMathews02}.  
\cite{pia03b} (hereafter MW03b) present constraints on $P_{in}(k)$
 from WMAP data using the wavelet band powers method, and also the 
tophat binning method for comparison. It was found that the reconstructed
 spectrum deviates from scale-invariant as well as power-law models by
 approximately 1$\sigma$. While the data do not rule out simple
 parametrizations
 such as scale-invariance, as also concluded by Bridle et al. (2003), 
Barger, Lee \& Marfatia (2003), Seljak, McDonald \& Makarov (2003), 
there is indication of excess power or a preferred scale at 
$k\sim 0.01$ Mpc$^{-1}$ (Peiris 
et al. 2003, Spergel et al. 2003, MW03b). 
\cite{WangMathews02}; Mukherjee \& Wang (2003a)
show that pre-MAP data indicated a similar feature.

It is worth investigating this issue further. Since data in CMB multipole 
space is essentially being used to constrain $P_{in}(k)$, its band powers
as estimated from the data become correlated with one another (because of
 the cosmological
 model dependent non-linear mapping
 between wavenumber $k$ and multipole $l$, see Tegmark \& Zaldarriaga 2002).
The $P_{in}(k)$ parameters in any model parametrization are correlated
 with the cosmological model parameters (that describe the evolution of
 structure from the initial conditions to the present). These correlations
 can make it tricky to efficiently constrain $P_{in}(k)$.

In this paper, we explore a more direct method of
 using wavelets
to probe $P_{in}(k)$; we parametrize $P_{in}(k)$ by
the coefficients of its wavelet expansion.
We obtain constraints on these wavelet coefficients and 
cosmological parameters from current CMB and LSS data.
This direct wavelet expansion method is different and complimentary to 
the wavelet band power method of MW03a,b.\footnote{The direct wavelet
expansion method uses a set of
 orthogonal basis functions that are very different from
those used in binning methods; the resultant constraints on 
$P_{in}(k)$ can be thus checked for consistency. This can be significant
 because parameter correlations,
which are important for this problem, are very different for the different
 sets of basis functions. We found the parameter correlations to be 
the smallest in 
 the direct wavelet expansion method.}  
It has the advantage that once the wavelet coefficients have 
been constrained, the reconstruction of $P_{in}(k)$ can be effectively
denoised, i.e., $P_{in}(k)$ can be reconstructed using only the 
coefficients that deviate from zero at greater than 1$\sigma$.
In doing so the essential properties of $P_{in}(k)$ are retained.
Because wavelets are adaptive, and the wavelet basis complete, any
 large scale as well as localized features in $P_{in}(k)$ can be efficiently
 picked up by a small number of coefficients.
Such properties make wavelets useful tools in signal reconstruction.  
This method is thus used to reconstruct $P_{in}(k)$ as an arbitrary function
 over the entire range in $k$ that cosmological data are sensitive
 to, and since it can be reliably reconstructed from a small number of 
wavelet coefficients, it suffers much less from the correlated 
errors of binning methods. 

We describe our method in Sec.2.  We present our results in Sec.3.
Sec.4 contains a summary and discussion.

\section{Method}

We expand the primordial power spectrum $P_{in}(k)$
\footnote{For a power-law primordial power spectrum,
$P_{\cal R}= 2.95 \times 10^{-9} \, A\,(k/k_0)^{n_S-1}$
\citep{Spergel03}. For an arbitrary primordial power spectrum,
we define 
$P_{in}(k)=\frac{P_{\cal R}}{2.95 \times 10^{-9}}$.}
directly in wavelet
basis functions,
\begin{equation}
P_{in}(k_i) = \sum_{j=0}^{J-1}\sum_{l=0}^{2^j-1} b_{j,l}\psi_{j,l}(k_i),
\label{WT}
\end{equation}
where $\psi_{j,l}$ are the wavelet basis functions,
constructed from the dilations and translations of a  
mother function $\psi(k)$, via
\citep{Daub92,Press94}
\begin{eqnarray}
\psi_{j,l}(k) & = & \left(\frac{2^j}{L}\right)^{1/2}\psi(2^jk/L-l).
\label{psibasis}
\end{eqnarray}
The resulting wavelet bases are discrete,
compactly supported and orthogonal with respect to both the scale $j$ 
and the position $l$ indices, and hence their coefficients provide a 
complete and non-redundant (hence invertible) representation of the
 function $P_{in}(k)$.
The scale index $j$ increases from 0 to $J-1$, and
 wavelet coefficients with increasing 
$j$ represent structure in $P_{in}(k)$ on
increasingly smaller scales, with each scale a factor of 2 finer than
the previous one.  The index $l$, which runs from 0 to $2^j-1$ for each $j$, 
denotes the position of the wavelet basis $\psi_{j,l}$ within the $j$th
scale. (Note that the properties of the basis functions are completely
 different from the basis functions in binning methods). Therefore the 
wavelet coefficients in Eq.(\ref{WT}), together with 
the scaling function coefficient which represents the mean of the function, 
(denoted $a_{0,0}$), 
 total to $2^J$ in number; correspondingly the function $P_{in}(k)$ is
 reconstructed at $2^J$ points.
 We have considered the case when these
 sample points $k_i$, over which the wavelet basis is defined and at
 which the function $P_{in}(k)$ is reconstructed, are equally spaced
 in $log(k)$. We use 16 wavelet coefficients to represent $P_{in}(k)$
in the range  $0.0002  \la k /(\mbox{Mpc}^{-1}) \la 0.2$ that the data
 are sensitive to, with the spectrum for $k<0.0002$ Mpc$^{-1}$ set to
 $P_{in}(k=0.0002\,$Mpc$^{-1}$) and the spectrum for $k>0.2$ Mpc$^{-1}$ set
 to $P_{in}(k=0.2\,$Mpc$^{-1}$).  

Thus, when the sample points 
are equally spaced, the coefficient $b_{j,l}$ measures the signal in 
$P_{in}(k)$ on scale $L/2^j$, 
and centered at $k=lL/2^j$, where $L$ is the full length 
over which $P_{in}(k)$ is being expanded.
The simultaneous and adaptive localization of the wavelet basis functions
 in position and frequency space make them useful tools for analyzing 
a variety of data.
 The information content of most signals can be efficiently encoded in 
wavelet space; hence the wavelet representation of a signal is often sparse
 and allows for significant compression (as coefficients below a threshold
 can be thrown away with very little loss of information making them 
important tools
 in denoising). For example a scale-invariant $P_{in}(k)$ requires only 
two non-zero coefficients (the scaling function coefficient, and the wavelet
 coefficient of scale 0), and when attempting to reconstruct this function
 from noisy data all other wavelet coefficients would be zero within error 
bars. If some other wavelet coefficients are non-zero then that implies 
additional structure in the function. 

Thus our goal is to constrain 
$P_{in}(k)$ by estimating its wavelet coefficients, together with 
cosmological parameters, from a likelihood analysis of cosmological data. 
We do not impose any particular inflation model dependent
 parametrization on $P_{in}(k)$, and allow
 it to be an arbitrary function. After constraining the wavelet 
coefficients, marginalizing over all other parameters considered,
 we can choose to reconstruct $P_{in}(k)$ using only those wavelet
 coefficients that are significantly constrained. In doing so all
 the important features of $P_{in}(k)$ are retained and one
 is essentially denoising. Thus in this construction features in 
any $P_{in}(k)$ can be localized
 in scale $j$ and position $l$ with the help of a few coefficients, and
 the resulting reconstruction also suffers much less from the correlated
 errors of binning methods.

This direct wavelet expansion method is very different in nature from the 
wavelet band power method used in MW03a. 
In the wavelet band power method of MW03a, the primordial power spectrum
is parametrized in terms of wavelet band powers; the window functions 
relating the wavelet band powers to the primordial power spectrum 
are the squares of the Fourier transforms of the wavelet basis functions (
apart from a $j$ dependent normalization factor). Wavelets that are more 
smooth and hence less localized in pixel space and correspondingly more
 localized in Fourier space, such as the Daubechies 20 wavelet are the suitable
 wavelets to use for this parametrization.
In the direct wavelet expansion method we use the coefficients of 
a wavelet expansion of $P_{in}(k)$ to characterize it;
 the window functions in this case are the mutually orthogonal 
wavelet basis functions $\psi_{j,l}$.   
Thus here we need to use wavelets that are most localized
in $k$ space, which is the pixel space here, such as the Haar and
 the Daubechies 4 (hereafter D4) wavelet. Among such wavelets different
 wavelets give very similar results. Other differences between the 
two methods are that in the wavelet band power method the bin locations 
and their separations are not arbitrary, as explained further in MW03a
(see also footnote 7 here), whereas here we can probe the function
 down to arbitrary resolution by reconstructing more coefficients.
The correlations among the reconstructed parameters are different
and smaller in the direct expansion method as compared to those in 
binning methods, and the reconstruction here can be effectively denoised;
these properties have been elaborated on earlier in this section.

We also make use of 
the mutual independence of the wavelet basis functions in computing 
the projections of the CMB angular power spectrum $C_l$'s and the matter 
power spectrum $P(k)$ for each wavelet basis, thus avoiding calling  
CMBFAST \citep{SeljakZ96} or CAMB \citep{Lewis00},
when only the 
wavelet coefficients are varied for the same cosmological parameters. 
This leads to a significant computational speed up. 
We write \footnote{In computing these 
wavelet projections, we note that $\psi_{k,l}$ are defined only at a
 finite number of points (here 16) and its value at any intermediate
 point is obtained by linear interpolation. This is appropriate; since we are
 using a discrete transform and correspondingly estimating a function
 at a finite number of points, no additional information at smaller
 scales is either assumed or included. We have checked that no
 interpolation errors are introduced at this stage.}

\begin{eqnarray}
C_l(\{b_{j,l}\}, \mbox{\bf s}) &=&(4\pi)^2 \int \frac{dk}{k} P_{in}(k) 
\left|\Delta_{Tl}(k, \tau=\tau_0)\right|^2
\nonumber\\
&=& \sum _{j,l} b_{j,l} \int \frac{dk}{k} \, \psi_{j,l}(k)
\left|\Delta_{Tl}(k, \tau=\tau_0)\right|^2\nonumber\\
&\equiv & \sum _{j,l} b_{j,l} \, C_l^{j,l} (\mbox{\bf s}),
\label{Clwaveletproj}  
\end{eqnarray}
where the cosmological model dependent transfer function 
$\Delta_{Tl}(k,\tau=\tau_0)$ is an integral over
 conformal time $\tau$ of the sources which generate CMB 
 temperature fluctuations,
 $\tau_0$ being the conformal time today,
 and $\mbox{\bf s}$ represents cosmological parameters other than
the $b_{j,l}$'s. 
The $C_l^{j,l}(\mbox{\bf s})$'s 
are the wavelet projections of the $C_l$'s, computed here 
using CAMB\footnote{http://camb.info/}. 

Similarly the wavelet projections
of the evolved matter power spectrum
$P(k)$ are computed for LSS data as follows,
\ba
P(k,\{b_{j,l}\},\mbox{\bf s}) & =&  P_{in}(k)*T(k)^2 \nonumber\\
       & = & \sum_{j,l} b_{j,l} \psi_{j,l}(k)*T(k)^2 \nonumber\\
       & \equiv & \sum_{j,l} b_{j,l} P^{j,l}(k,\mbox{\bf s})
\ea      
where $T(k)$ is the matter transfer function.

Since the theoretical predictions to be compared with CMB and 
LSS data both depend on $P_{in}(k)$, and the cosmological parameters, 
they can together
 be used to place robust constraints on cosmological models and their
 initial conditions. Note however that as pointed out by Elgaroy,
 Gramann, \& Lahav (2002), it is hard to detect features in the
 primordial power spectrum at $k<0.03h$Mpc$^{-1}$ using LSS data only.

We use the Markov Chain Monte Carlo (MCMC) technique,
illustrated for example in \cite{LB02,Verde03} to estimate the 
likelihood functions of the parameters.
This technique is necessitated by the large number 
of parameters being varied here, 
and at the same time is free of the interpolation errors expected in the
 much slower grid based methods.
At its best, the MCMC method scales approximately linearly with the number of 
parameters.  See \cite{neil} for a review, and
 \cite{hannestad, Knox01, Rubino-Martin02} for other applications of
 this method to CMB data analysis.  An MCMC sampler 
 based on the Metropolis-Hastings
algorithm has been made 
 available in the software package CosmoMC \citep{LB02}.

The MCMC method is used to trace the full posterior distribution of
 the parameters in a flat $\Lambda$CDM cosmology assuming wide uniform 
priors of $0.4<h<1.0$, where
 $h = H_0/$(100 kms$^{-1}$ Mpc$^{-1}$, $0.005< \Omega_b h^2 < 0.1$
 and $0.1< \Omega_c h^2 <0.99$ on the cosmological parameters. The
 two wavelet coefficients that set the level of $P_{in}(k)$ are allowed
 to vary between 0 and 5, while the remaining wavelet coefficients are
 allowed to vary between $\pm 2.5$, with the requirement that the 
resulting $P_{in}(k)$ be positive definite\footnote{Note that for a 
scale-invariant $P_{in}(k)$ of unity the two coefficients that set 
the level are each 2.828 while the remaining 14 coefficients are 0.}.
 We use a weak prior on
 the age of the universe of $t_0 > 10$ Gyrs. We have run several chains, 
 and have checked for convergence and mixing. From the MCMC samples thus 
obtained, the marginalized posterior distributions and confidence limits
 of the parameters are estimated.

\section{Results}

We use CMB temperature anisotropy data from WMAP (Bennett et al. 2003),
 complimented at $l>800$ by data from CBI (Pearson et al. 2002) and ACBAR
 (Kuo et al. 2002) upto an $l_{max}$ of 2000, marginalizing analytically 
over known beamwidth and calibration uncertainties (Bridle et al. 2002).
 We use LSS data from 2dFGRS 
(Percival et al. 2002; the 
redshift space power spectrum is assumed to be proportional to the real 
space power spectrum on the large scales used) and PSCZ (Hamilton \& 
Tegmark 2002; decorrelated real space power spectrum) galaxy redshift 
surveys over linear scales. A linear bias is assumed and analytically
 marginalized over (Bridle et al. 2002). LSS data are thus effectively
 being used to constrain the shape and not directly the amplitude of
 the matter power spectrum.

We assume Gaussian adiabatic scalar perturbations in a flat universe 
with a cosmological constant. We do not use tensor modes in this paper,
 since current data are not sensitive to tensor contributions. With
 freedom allowed in the shape of the primordial power spectrum, the
 temperature anisotropy data themselves cannot distinguish well
 between different values of the reionization optical depth $\tau_{ri}$.
 Hence we simply make use of the WMAP constraint of $\tau_{ri}=0.17\pm0.04$,
 derived for a flat $\Lambda$CDM cosmology from WMAP's TE polarization data 
\citep{Kogut03}, where the error estimate includes systematic
 and foreground uncertainties. We marginalize the distributions of the
 other parameters over a Gaussian distribution in $\tau_{ri}$ centered at
 0.17 with a standard deviation of 0.04. For this, we calculate the likelihood
 distributions of the wavelet coefficients as well as the Hubble 
constant $H_0$, baryon density $\Omega_b\,h^2$ and the cold dark matter
 density $\Omega_c\,h^2$ at 5 select values of $\tau_{ri}$ and approximate
 the marginalization over $\tau_{ri}$ using the 5-point Gauss-Hermite
 quadrature summation. 
 
Fig.1 shows the marginalized 1d distributions of the parameters, obtained
 from an MCMC based likelihood analysis of CMB and LSS data, after 
marginalizing over the assumed distribution for $\tau_{ri}$.
 All the parameters are well constrained, 
lying well within the imposed priors. 

Fig.2 shows the reconstructed $P_{in}(k)$ (solid curves) and its
 1$\sigma$ confidence region (dotted curves), obtained from the mean 
 values and standard deviations of the wavelet
 coefficients.  The uncertainty in $P_{in}(k)$ has been
 calculated using:
\begin{equation}
\sigma^2_{P_{in}(k_i)} = \frac{1}{N} \sum_N \left[ \sum_{j,l}
 (b_{j,l}-\bar{b}_{j,l}) \frac{\partial P_{in}(k_i)}{\partial b_{j,l}} 
\right]^2,
\end{equation}   
where the average is over the MCMC samples, $\bar{b}_{j,l}$ denotes 
the mean value of the wavelet coefficient, and $\frac{\partial
 P_{in}(k_i)}{\partial b_{j,l}}=\psi_{j,l}(k_i)$; correlations between 
the wavelet coefficients have thus been accounted for in computing
 the uncertainty in $P_{in}(k)$.
The upper panel shows the reconstructed $P_{in}(k)$ 
 using all wavelet coefficients. These constraints are shown overlaid on the 
constraints obtained by MW03b using the wavelet band power 
method.\footnote{Note that in the wavelet banding 
method the locations and the widths of the bands are not arbitrary. As
 explained in Mukherjee \& Wang 2003a, the band power window functions
 in this case are the modulus squared of the Fourier transforms of the
 wavelet basis functions. By definition then, since each scale is a factor
 of 2 finer than the previous one, 
 the bands lie $log_{10}2$ apart in $k$.}
We see 
 that while the constraints from the two methods are completely
 consistent, a new feature shows up at large angular scales
 around $0.001 \la k/\mbox{Mpc}^{-1} \la 0.005$. 
 The lower panel shows the spectrum 
 reconstructed using only
 those coefficients that deviate from 0 at $>1\sigma$.
These constraints are shown overlaid on the constraints obtained using
 the tophat binning method with bands carefully positioned closer together 
 so as to be able to pick up the feature at large angular scales
 around $0.001 \la k/\mbox{Mpc}^{-1} \la 0.005$, and placed further 
 apart in the remaining
parts of the spectrum.
  The feature shows up consistently.

The coefficients that are significant at the greater than 1$\sigma$ level 
are $b_{0,0}$, $b_{1,0}$, $b_{2,0}$, $b_{2,1}$ and $b_{3,2}$, together
 with the scaling function coefficient. Notice that with 10 of the 16
 wavelet coefficients dropped, the spectrum reconstructed in panel (b)
 still has all the essential features of the spectrum shown in panel (a).
The rise in power that shows up near $k\sim 0.01$ Mpc$^{-1}$ mimics a
 power-law model with possible running and is 
what was reported in MW03b and by WMAP. In this paper we report a
 localized new feature at $0.001 \la k/\mbox{Mpc}^{-1} \la 0.005$
 that would be missed by a power-law parametrization,
 and by other model-dependent parametrizations.
(We note that section
 5 of Peiris et al. (2003) seems to indicate the same large scale feature
 obtained here, but they used the model dependent parametrization of
 Adams, Cresswell, \& Easther (2001) and constrained the $P_{in}(k)$ parameters
 at a fixed set of cosmological parameters.) This new feature would
 also be missed by a model-independent binning 
 method if the bin separation were not small enough
 at the appropriate locations.\footnote{Note that a uniform binning
 with a small bin size can lead to oversampling of the data, which
 would result in large correlated error bars for the estimated bin amplitudes,
 thus burying any possible features (Bridle et al. 2003).} 

In the direct expansion method the constraints obtained 
simultaneously on the cosmological parameters are 
$h=0.59\pm0.09$, $\Omega_b h^2=0.018\pm 0.004$ and 
$\Omega_c h^2 = 0.143\pm 0.019$. In all the constraints presented here we
 have used the parameter mean values together with marginalized 
1$\sigma$ standard deviations.\footnote{As discussed
 in \cite{LB02}, the MCMC samples from the posterior do not provide accurate
 estimates of parameter best-fit values, because in higher dimensions the 
best-fit region typically has a much higher likelihood than the mean but
 occupies a very small fraction of parameter space.}
 
The $\chi^2_{eff}$ of the best fit model in the direct wavelet 
expansion method is 1032.45 for a total of 964 data points, when
 constraining 16 wavelet coefficients and 3 cosmological
 parameters with an external prior
 on $\tau_{ri}$ being imposed and marginalized over.\footnote{The number 
 of data points in the
 different data sets used are 899 (WMAP), 4 (CBI), 7 (ACBAR), 32
 (2dFGRS) and 22 (PSCZ).} 
 A scale-invariant 
model with 5 constrained parameters has a $\chi^2_{eff}$ of 1044.33 
for the same data. One can see that this model is equivalent to 
holding fixed the 14 wavelet coefficients that are responsible 
for any deviations from scale-invariance. Thus the difference in
 $\chi^2$ between the two models has a $\chi^2$-distribution with
 14 degrees of freedom. By computing the probability of obtaining such
 a $\chi^2$ we are able to compare the 
 two models and find that the scale-invariant
 model is disfavoured at $\sim\,0.6\sigma$. The $\chi^2_{eff}$ of the
 best fit model in the direct wavelet expansion method using only
 the 6 significantly constrained wavelet coefficients is 1039.52
 for 9 constrained parameters. This is favored over the scale-invariant
 model at $\ga\, 1\sigma$.\footnote{The $\chi^2_{eff}$ of the best fit
 power-law model for these data is 1043.68 with 6 constrained parameters.
 In order to compare this model with the wavelet coefficients based model
 one would need to use a more involved method, possibly that of 
Bayesian evidence.}
Note that in our analysis the number
 of constrained parameters are large and the significance of deviations
 from scale-invariance are small because we allow $P_{in}(k)$ to be an 
arbitrary function over the entire $k$ range. While the
 deviation in the shape
 of $P_{in}(k)$ from scale-invariance is not found to be
 significant at this stage, 
 the primordial power spectrum
reconstructed here seems to indicate 
 intriguing features that could indicate either unusual inflationary
 physics or hidden systematics in current CMB data.

\section{Conclusions}
It is important to reconstruct the primordial power spectrum as a free function
from cosmological data, the aim being 
 to test the assumptions that are usually made about $P_{in}(k)$,
 and to constrain it model-independently so as to be able
 to detect any features in it that may be signatures of unusual physics 
during inflation.
We have presented a new and powerful method to extract
the primordial power spectrum $P_{in}(k)$ as an arbitrary function from 
observational data --- the direct expansion of $P_{in}(k)$ 
in wavelet basis functions.

The wavelet basis functions afford an efficient (sparse)
and non-redundant (invertible) expansion of a function,
with coefficients that encode information about the structure of 
the function on different scales, 
and with adaptive position resolution. These properties make
wavelets suitable tools for directly probing the primordial power 
spectrum for features. Reconstructing $P_{in}(k)$ using only the 
wavelet coefficients 
that deviate from 0 at $>1\sigma$ amounts to denoising the
 resultant power spectrum while retaining all its important features (Fig 2).
The direct wavelet expansion method also yields smaller parameter correlations
than the binning methods.

The $P_{in}(k)$ reconstructed in this paper (Fig.2) is favored over the
scale-invariant and power-law parametrizations at $\ga 1\sigma$.
The results of this paper are completely
 consistent with 
the spectrum reconstructed in MW03b using the wavelet band power method.
These two methods are very different in nature and complementary 
 (see section 2).

The direct wavelet expansion method has enabled us to  
find indication of a new localized feature in $P_{in}(k)$
at large scales around
$0.001 \la k/\mbox{Mpc}^{-1} \la 0.005$.
It will be interesting to see whether future data will confirm the
existence of this feature.

\acknowledgements
We acknowledge helpful discussions with Antony Lewis, 
Christopher Koenigsberg, and Rich Holman. 
We thank Henry Neeman for computational support.
We acknowledge the use of CAMB and CosmoMC. This work is
supported in part by NSF CAREER grant AST-0094335. 





\begin{figure}
\plotone{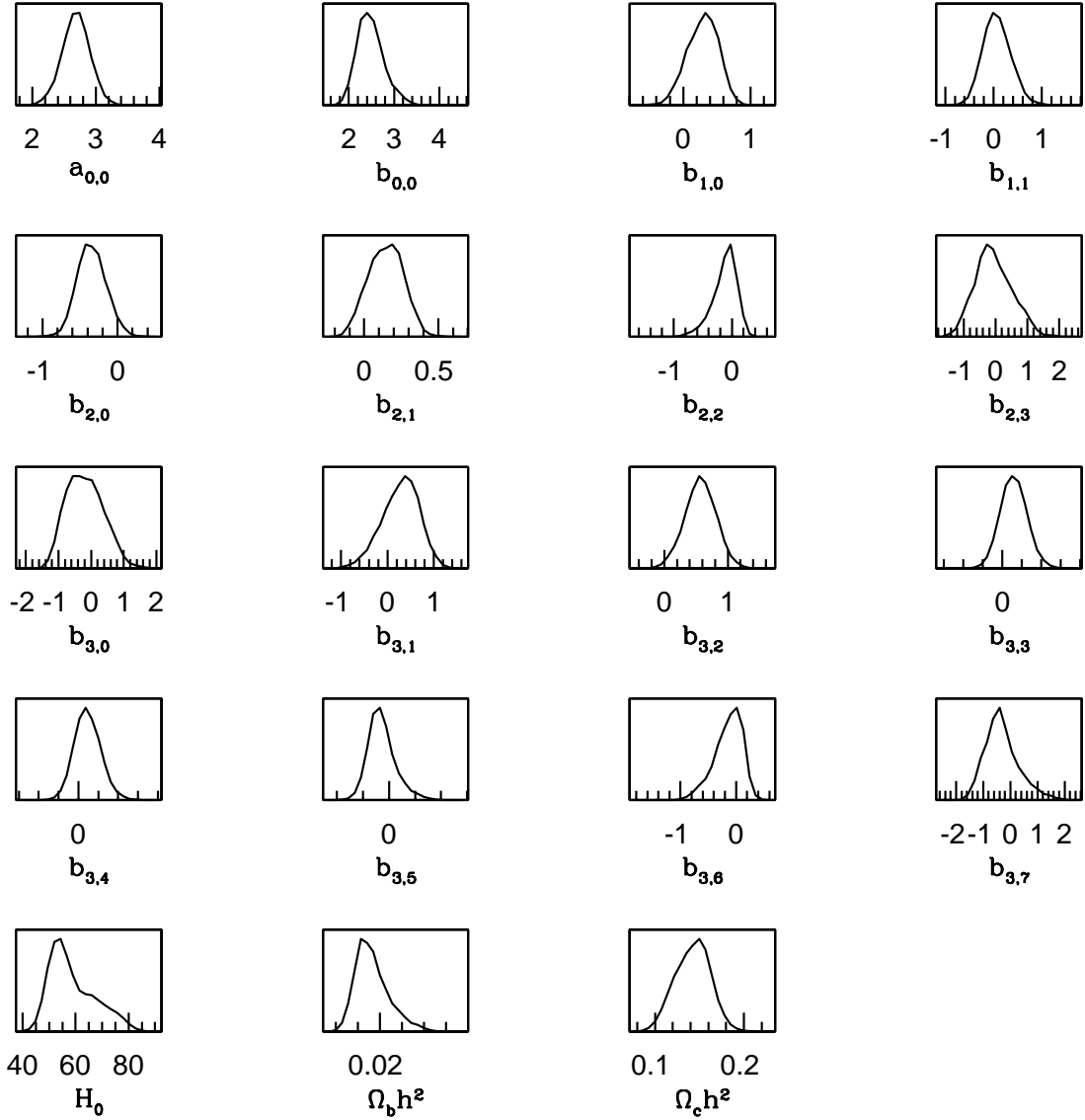}
\figcaption{
The figure shows the marginalized 1d distributions of all the
 parameters used; 16 wavelet coefficients and 3 cosmological parameters.
 A Gaussian distribution of $\tau$ has been marginalized over (see text).
 The mean values and 1$\sigma$ standard deviations on the cosmological
 parameters are $h=0.59\pm0.09$, $\Omega_b h^2=0.018\pm 0.004$ and 
$\Omega_c h^2 = 0.143\pm 0.019$. Corresponding constraints on the 
wavelet coefficients imply constraints on $P_{in}(k)$ shown in Figure 2.
}
\end{figure}




\begin{figure}
\plotone{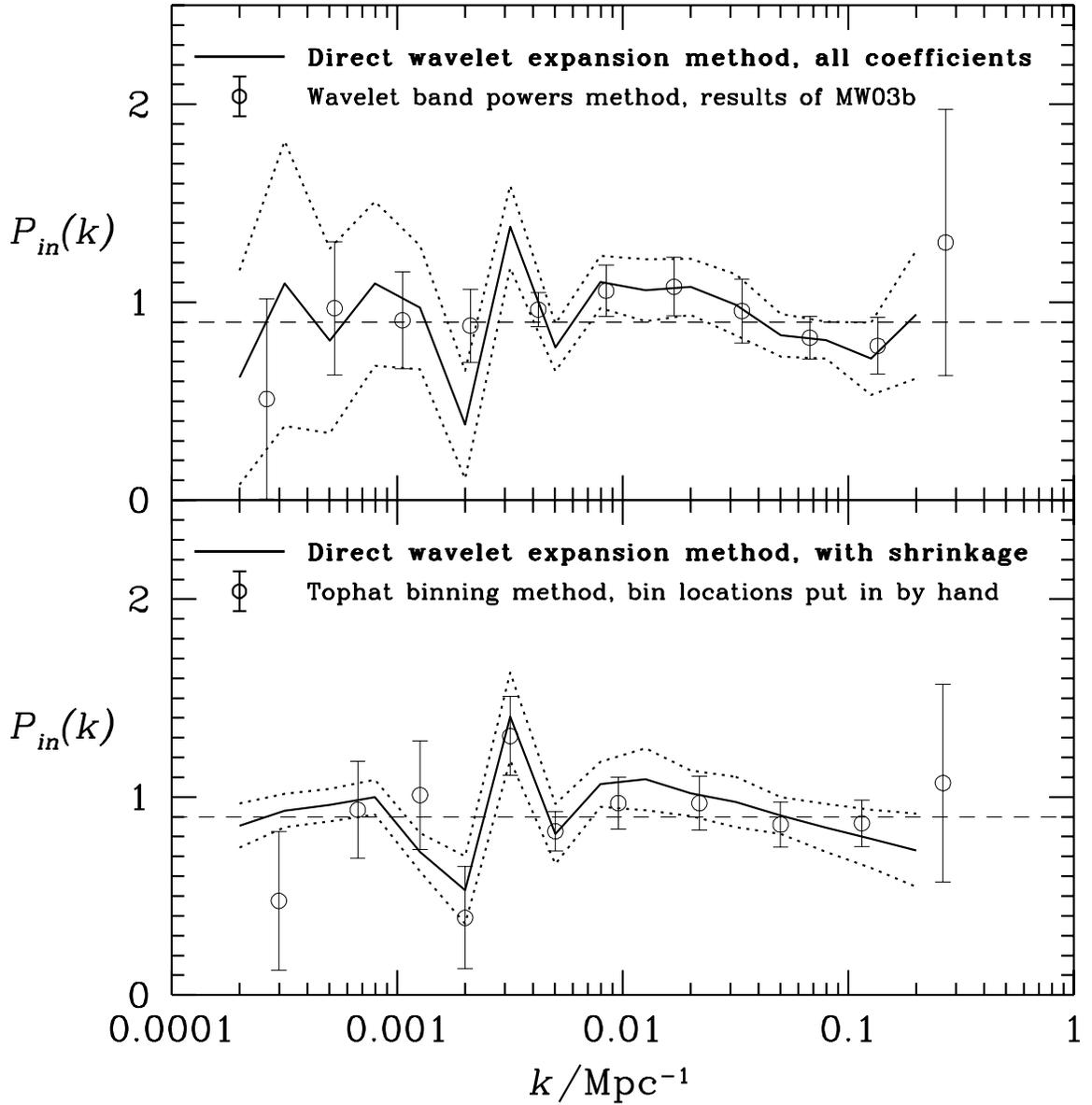} 
\figcaption{
reconstructed using all wavelet coefficients (solid curve) with
 its 1$\sigma$ confidence region (dotted curves). 
These constraints are shown overlaid on the 
constraints obtained by MW03b using the wavelet band power method.
(b) $P_{in}(k)$ reconstructed using only
 those coefficients that deviate from 0 at $>1\sigma$ (solid curve) with
 the corresponding 1$\sigma$ confidence region (dotted curves).
These constraints are shown overlaid on the constraints obtained using
 the tophat binning method with bands carefully positioned closer together 
 so as to be able to pick up the feature at large angular scales
 around $0.001 \la k/\mbox{Mpc}^{-1} \la 0.005$, and placed further 
 apart in the remaining
 parts of the spectrum. The dashed line is the scale-invariant
 spectrum that best fits these data.
}

\end{figure}




\end{document}